\documentstyle[epsfig,11pt]{article}

\newlength{\dinwidth}
\newlength{\dinmargin}
\newcommand{\ba}{\begin{array}}
\newcommand{\ea}{\end{array}}
\newcommand{\bd}{\begin{displaymath}}
\newcommand{\ed}{\end{displaymath}}
\newcommand{\be}{\begin{equation}}
\newcommand{\ee}{\end{equation}}
\newcommand{\bea}{\begin{eqnarray}}
\newcommand{\eea}{\end{eqnarray}}
\def\a{\alpha}
\def\b{\beta}

\def\G{\Gamma}

\begin{document}

\bigskip

\title{Study of $B\to K^* \rho,~K^*\omega$ Decays with
Polarization in Perturbative QCD Approach }

\author{Han-Wen Huang$^{a,b}$,
Cai-Dian L\"u$^{c,a}$, Toshiyuki Morii$^d$, Yue-Long
Shen$^a{\footnote {shenyl@mail.ihep.ac.cn}}$, Ge-Liang Song$^e$,
and Jin Zhu$^a$
 \\
{\it \small $a$ Institute of High Energy Physics,
 P.O. Box 918(4), Beijing 100049, China}\\
{\it \small $b$ Department of Physics, University of Colorado,
Boulder, CO80302, USA}\\
{\it \small $c$ CCAST (World Laboratory),
 P.O. Box 8730, Beijing 100080, China}\\
{\it \small $d$ Faculty of Human Development, Kobe University,
Nada, Kobe 657-8501, Japan}\\
{\it \small $e$ Dept. of Physics, University of Notre Dame du Lac,
Notre Dame, IN 46556, USA}}

\maketitle

\begin{abstract}

The $B  \to K^{*}\rho$, $ K^{*}\omega$ decays are useful to
determine the CKM angle $\phi_3=\gamma$. Their  polarization
fractions are also interesting since the polarization puzzle of
the $B\to \phi K^*$ decay. We study these decays in the
perturbative QCD approach based on $k_T$ factorization.
 After calculating of the
non-factorizable and annihilation type contributions, in addition
to the conventional factorizable contributions, we find that the
contributions from the annihilation diagrams are crucial.  They
give dominant contribution to the strong phases and suppress the
longitudinal polarizations. Our results agree with the current
existing data. We also predict a sizable direct CP asymmetries in
$B^+ \to K^{*+}\rho^0$, $B^0 \to K^{*+}\rho^-$, and $B^+ \to
K^{*+}\omega$ decays,   which  can be tested by the oncoming
measurements in the  B factory experiments.
\end{abstract}

\bigskip


\newpage

\section{Introduction}

The hadronic $B$ decays have been studied for many years since
they offer an excellent  place to study the CP violation and
search for new physics hints \cite{sanda}. The hadronization of
the final states is nonperturbative in nature, and the essential
problem in handling the decay processes is the separation of
different energy scales, namely, the so-called factorization
assumption. Many factorization approaches have been developed  to
calculate the $B$ meson decays, such as the naive factorization
\cite{fac}, the generalized factorization  \cite{akl1,cheng}, the
QCD factorization  \cite{bene}, as well as the perturbative QCD
approach (PQCD) based on $k_T$ factorization \cite{kls,lucd}. Most
factorization approaches are based on heavy quark expansion and
light-cone expansion,   only the leading power or part of next to
leading power contributions are calculated to compare with the
experiments. Nevertheless for the penguin-dominated decay
channels, the power corrections and the nonperturbative
contributions may be large, since the theoretical predictions for
some channels cannot fit the data quite well. There are   some
  problems, such as the $\sin{2\beta}$ problem in penguin dominated modes \cite{2beta},
which suggests   that more dynamics of penguin dominating B decays
should be studied.

 Recently, with more and more data, the B factories have measured
 some decays that the final state contains two vector mesons \cite{kphi,rhorho,babar}.
  In the $B\to VV$ modes, both the longitudinal and the transverse polarization can
 contribute to the decay width, and the polarization fractions can be
 measured by the experiments. The naive counting rules based on the
 factorization approaches predict
 that the longitudinal polarization dominates the decay ratios and
 the transverse polarizations are suppressed \cite{power} due to the helicity
 flips of the quark in the final state hadrons.
 But some data shown in table 1 are quite different from the theoretical
 predictions for the penguin dominated modes.

 \begin{table}\begin{center}\caption{Longitudinal Polarization fractions of some $B\to VV$ modes. }
\begin{tabular}{|l|l|l|l|l|l|}
\hline
 Process & Belle &Babar & QCDF
 \cite{yang,chengyang}&QCDF+FSI\cite{fsi}
  \\ \hline
 $ B^0\to\phi K^{*0 }$&$0.45\pm 0.05 \pm 0.02 $& $0.52\pm 0.05 \pm 0.02$
   & 0.91&$0.43^{+0.13}_{-0.09}$
\\ \hline
 $ B^+\to\phi K^{*+ }$&$0.52\pm 0.8 \pm 0.03 $& $0.46\pm 0.12 \pm 0.03$
   & 0.91&$0.43^{+0.13}_{-0.09}$
\\ \hline
 $ B^+\to\rho^0 K^{*+ }$&  & $0.96^{+0.04}_{-0.15}\pm 0.04 $
   & 0.94&$0.49^{+0.11}_{-0.08}$

\\ \hline
 $ B^+\to\rho^+ K^{*0 }$&$0.43\pm 0.11^{+0.05}_{-0.07} $& $0.79\pm 0.08 \pm 0.04\pm 0.02$
   & 0.95&$0.57^{+0.16}_{-0.14}$
\\ \hline

\end{tabular}
\end{center}\end{table}

   The     small   longitudinal
polarization fraction in $B\to \phi K^{\ast}$ decays has been
considered as a puzzle,   many theoretical efforts have been
performed to explain it
 \cite{grossman,yang,kagan,scet,fsi,hou,lm,yangkc}.
 In PQCD approach, the coefficients of penguin
operators have been evolved to the scale of about $\sqrt{\Lambda
M_B}$, so these coefficients become larger compared to the
factorization approach, in which the hard scale are at the scale
of $M_B$, so that the penguins' contribution are enhanced in PQCD
approach. Besides, the annihilation diagrams, which is power
suppressed in QCD factorization, are also included. Thus   the
PQCD approach can give a larger branching ratio and fits the
experiments  well in   $B\to PP, PV$ case. For $B\to \phi
K^{\ast}$, the annihilation diagram with the $(S+P)(S-P)$ type
operators will break the naive counting rules \cite{kagan}, the
transverse polarization is enhanced to about $0.25$. But the
branching ratios calculated in the PQCD approach \cite{kphipqcd}
are too large if we adopt the old $K^{\ast}$ meson's parton
distribution amplitudes derived from QCD sum rules. As mentioned
in
 \cite{reso}, things will get better (59\% of longitudinal polarization)
 if we adopt the asymptotic
form of the $K^{\ast}$ meson's parton distribution amplitudes.

In this paper, we will perform the leading order PQCD calculation
of penguin dominated processes $B\to \rho K^{\ast}$ and $B\to
\omega K^{\ast}$. The branching ratios have been measured by the B
factories \cite{hfag} which are given in table 2. And the measured
CP asymmetries are: ${\cal A}_{CP}( B^+\to\rho^+ K^{*0 })=-0.14\pm
0.17 \pm 0.4$ and ${\cal A}_{CP}( B^+\to\rho^0 K^{*+
})=0.20^{+0.32}_{-0.29} \pm 0.04$. These channels have been
studied within the QCD factorization framework, but the
predictions are not quite consistent with the data, especially the
polarization fractions \cite{chengyang}. We hope the PQCD approach
could give a better theoretical prediction.

\begin{table}
\begin{center}\caption{  Branching ratios ($10^{-6}$) of $B\to \rho K^{\ast}$
  measured by the B factories}
\begin{tabular}{|l|l|l|l|l|l|}
\hline
 Process & BaBar &Belle & world average
  \\ \hline
 $ B^0\to\rho^- K^{*+ }$&$< 24$&
   & $< 24$
\\ \hline
 $ B^+\to\rho^0 K^{*+ }$&$10.6^{+3.0}_{-2.6}\pm 2.4$&
   & $10.6^{+3.8}_{-3.5}$
\\ \hline
 $ B^+\to\rho^+ K^{*0 }$& $17.0\pm 2.9^{+2.0}_{-2.8}$ & $8.9\pm 1.7\pm 1.0 $
   & $10.5\pm 1.8$
\\ \hline
 $ B^0\to\rho^0 K^{*0 }$& & $<2.6$
   & $<2.6$
\\ \hline
 $ B^+\to\omega K^{*+ }$&$<7.4$&
   & $<7.4$
\\ \hline
   $B^0\to\omega K^{*0 }$&$<6.0$&
   & $<6.0$
\\ \hline
\end{tabular}
\end{center}\end{table}

The paper is organized as follows: In Sect. 2 we will present the
framework for three scale PQCD factorization theorem. Next we will
give the perturbative calculation result for the hard part. In
Sect. 4, numerical calculation for branching ratio and CP
violation are given. Final section is devoted to summary.

\section{The Theoretical Framework}

The  PQCD factorization theorem has been developed for
non-leptonic heavy meson decays  \cite{li}, based on the formalism
by Brodsky and Lepage  \cite{bl}, and Botts and Sterman
\cite{bs}. In the two body hadronic $B$ decays, the $B$ meson is
heavy, sitting at rest. It decays into two light mesons with large
momenta. Therefore the light mesons are moving very fast in the
rest frame of $B$ meson.

 To form the fast moving final state light meson, in which the two
 valence quarks  should be collinear, there must be a hard gluon to
 kick off the light spectator quark
$d$ or $u$ in the $B$ meson (at rest). So the contribution from
the hard gluon exchange between the spectator quark and the quarks
which form the four quark operator dominates the matrix element of
the four quark operator between hadron states. This process can be
calculated perturbatively, but the endpoint singularity will
appear if we drop the transverse momentum carried by the quarks.
After introducing the parton's transverse momentum, the
singularity is regularized, and additional energy scale is present
in the theory, then the perturbative calculation will produce
large double logarithm terms, these terms are then resummed to the
 Sudakov form factor. The uncancelled soft and collinear
divergence should be absorbed into the definition of the meson's
wave functions, then the decay amplitude is infrared safe and can
be factorized as the following formalism:

\begin{equation}
C(t) \times H(t) \times \Phi (x) \times \exp\left[ -s(P,b)
-2 \int _{1/b}^t \frac{ d \bar\mu}{\bar \mu} \gamma_q (\alpha_s (\bar \mu))
\right], \label{eq:factorization_formula}
\end{equation}
where $C(t)$ are the corresponding Wilson coefficients of four
quark operators, $\Phi (x)$ are the meson wave functions and the
variable $t$ denotes the largest energy scale of hard process $H$,
it is the typical energy scale in PQCD approach and the Wilson
coefficients are evolved to this scale.  The exponential of $S$
function is the so-called Sudakov form factors, which can suppress
the contribution from the nonperturbative region, making the
perturbative region give the dominated contribution. The
``$\times$'' here denotes convolution, i.e., the integral on the
momentum fractions and the transverse intervals of the
corresponding mesons. Since logarithm corrections have been summed
by renormalization group equations, the factorization above
formula does not depend on the renormalization scale $\mu$
explicitly.

In the resummation procedures, the $B$ meson is treated as a
heavy-light system. In general, the $B$ meson light-cone matrix
element can be decomposed as  \cite{grozin,bene}
\begin{eqnarray}
&&\int_0^1\frac{d^4z}{(2\pi)^4}e^{i\bf{k_1}\cdot z}
  \langle 0|\bar{b}_\alpha(0)d_\beta(z)|B(p_B)\rangle \nonumber\\
&=&-\frac{i}{\sqrt{2N_c}}\left\{(\not p_B+m_B)\gamma_5
\left[\phi_B ({\bf k_1})-\frac{\not n_+-\not n_-}{\sqrt{2}}
\bar{\phi}_B({\bf k_1})\right]\right\}_{\beta\alpha}, \label{aa1}
\end{eqnarray}
where $n_+=(1,0,{\bf 0_T})$, and $n_-=(0,1,{\bf 0_T})$ are the
unit vectors pointing to the plus and minus directions,
respectively. As pointed out in ref.\cite{collins}, this kind of
definition will provide light-cone divergence, and more involved
studies have been performed \cite{liao,ma}. Here we only use it
phenomenologically to fit the data, so we still use the old form.
 From the above equation, one can see that there are two Lorentz
structures in the $B$ meson distribution amplitudes. They obey the
following normalization conditions
\begin{equation}
\int \frac{d^4 k_1}{(2\pi)^4}\phi_B({\bf
k_1})=\frac{f_B}{2\sqrt{2N_c}}, ~~~\int \frac{d^4
k_1}{(2\pi)^4}\bar{\phi}_B({\bf k_1})=0.
\end{equation}
In general, one should consider both of these two Lorentz
structures in calculations of $B$ meson decays. However, it can be
argued that the contribution of $\bar{\phi}_B$ is numerically
small  \cite{kurimoto,form}, thus its contribution can be
 neglected. Therefore, we only consider the
contribution of Lorentz structure
\begin{equation}
\Phi_B= \frac{1}{\sqrt{2N_c}} (\not\! p_B +m_B) \gamma_5 \phi_B
({\bf k_1}), \label{bmeson}
\end{equation}
in our calculation. Note that we use the same distribution
function $\phi_B (k_1)$ for the $\not \! p_B$ term and the $m_B$
term in the heavy quark limit. For the hard part calculations in
the next section, we use the approximation $m_b\simeq m_B$, which
is the same order approximation neglecting higher twist of
$(m_B-m_b)/m_B$. Throughout this paper, we take light-cone
coordinates, then the four momentum $p^\pm = (p^0 \pm
p^3)/\sqrt{2}$, and ${\bf p}_T = (p^1, p^2)$.  We consider the $B$
meson at rest, the momentum is $p_B=(m_B/\sqrt{2}) (1,1,{\bf
0}_T)$. The momentum of the light valence quark is written as
($k_1^+,k_1^-, {\bf k}_{1T}$), where the $ {\bf k}_{1T}$ is a
small transverse momentum. It is difficult to define the function
$\phi_B(k_1^+,k_1^-,{\bf k}_{1T})$. However, the hard part isn't
always dependent on $k_1^+$ if we make some approximations. This
means that $k_1^+$ can be simply integrated out for the function
$\phi_B(k_1^+,k_1^-,{\bf k}_{1T})$ as
\begin{eqnarray}
\phi_B (x, {\bf k}_{1T}) &=&\int d k_1^+ \phi_B (k_1^+, k_1^-, {\bf k}_{1T})
\label{int}
\end{eqnarray}
where $x=k_1^-/p_B^-$ is the momentum fraction. Therefore, in the
perturbative calculations, we do not need the information of all
four momentum $k_1$. The integration above can be done only when
the hard part of the subprocess is independent on the variable
$k_1^+$.

The $K^*$ and $\rho$ mesons are treated as a light-light system.
At the $B$ meson rest frame, they are moving very fast. We define
the momentum of the $K^*$ as $P_2=(m_B/\sqrt{2})(1-r_3^2,r_2^2,
{\bf 0}_T)$. The $\rho$ has momentum $P_3= (m_B/\sqrt{2})
(r_3^2,1-r_2^2, {\bf 0}_T)$, with $r_2=M_{K^*}/M_B$ and
$r_3=M_{\rho(\omega)}/M_B$.  The light spectator quark in $K^*$
meson has a momentum $(k_2^+, 0, {\bf k}_{2T})$. The momentum of
the other valence quark in this final meson is thus $(P_2^+ -
k_2^+, 0, -{\bf k}_{2T})$. The longitudinal polarization vectors
of the $K^*$ and $\rho$ are given as:
\begin{eqnarray}
\epsilon_2(L)=\frac{P_2}{M_{K^*}}-\frac{M_{K^*}}{P_2\cdot n_-}n_-\;,
\epsilon_3(L)=\frac{P_3}{M_\rho}-\frac{M_\rho}{P_3\cdot n_+}n_+\;,\;\;\;\;
\end{eqnarray}
which satisfy the normalization
$\epsilon_2^2(L)=\epsilon_3^2(L)=-1$ and the orthogonality
$\epsilon_2(L)\cdot P_2=\epsilon_3(L)\cdot P_3=0$ for the on-shell
conditions $P_2^2=M_{K^*}^2$ and $P_3^2=M_\rho^2$. We first keep
the full dependence on the light meson masses $M_{K^*}$ and
$M_\rho$ with the momenta $P_2$ and $P_3$. After deriving the
factorization formulas, which are well-defined in the limit
$M_{K^*},M_{\rho}\to 0$, we drop the terms proportional to
$r^{2}_{\rho},\ r_{K^*}^2\sim 0.04$. The transverse polarization
vectors can be adapted directly as
\begin{eqnarray}
\epsilon(+) =\frac{1}{\sqrt{2}}(0,0,1,i)\;,\;\;\;\;
\epsilon(-) =\frac{1}{\sqrt{2}}(0,0,1,-i)\;.
\end{eqnarray}

If the ${K^*}$ meson (so as to other vector mesons) is longitudinally
polarized, we can write its wave function in longitudinal polarization
 \cite{kurimoto,ball2}
\begin{eqnarray}
&&<{K^*}^-(P,\epsilon_L)|\overline{s}_{\alpha}(z)u_{\beta}(0)|0>\nonumber\\
&=& \frac{1}{\sqrt{2N_c}}\int_0^1 dx e^{ixP\cdot z} \left\{ \not\!
\epsilon \left[\not \!p_{K^*} \phi_{K^*}^t (x) + m_{K^*}
\phi_{K^*} (x) \right] +m_{K^*} \phi_{K^*}^s (x)\right\} .
\end{eqnarray}
The second term in the above equation is the leading twist wave
function (twist-2), while the first and third terms are sub-leading
twist (twist-3) wave functions. If the $K^*$ meson is transversely
polarized, its wave function is then
\begin{eqnarray}
<{K^*}^-(P,\epsilon_T)|\overline{s}_{\alpha}(z)u_{\beta}(0)|0> &=&
\frac{1}{\sqrt{2N_c}}\int_0^1 dx e^{ixP\cdot z} \left\{\not\!
{\epsilon} \left[\not\! p_{K^*}
\phi_{K^*}^T (x) + m_{K^*} \phi_{K^*}^v (x) \right] \nonumber\right.\\
&&\left.+ i m_{K^*} \epsilon_{\mu\nu\rho\sigma}\gamma_5\gamma^\mu
\epsilon^\nu n^\rho v^\sigma\phi_{K^*}^a (x)\right\} .
\end{eqnarray}
 Here the leading twist wave function for the transversely polarized
 $K^*$ meson is the first term which is proportional to $\phi_{K^*}^T
 $.

The transverse momentum ${\bf k}_{iT}$ is usually converted to the
$b$ parameter by Fourier transformation. The initial conditions of
$\phi_i(x), i=B,K^*,\rho$, are of nonperturbative origin,
satisfying the normalization
\begin{equation}
\int_0^1\phi_i(x,b=0)dx=\frac{1}{2\sqrt{2N_c}}f_i, \label{eq:norm}
\end{equation}
with $f_i$ the meson decay constants.

\section{Perturbative Calculations}
\label{sc:PertCalc}

With the above brief discussion, the only thing left is to compute
the hard part $H$ . We use the notation
$M_\lambda=<V_1(\lambda)V_2(\lambda)|H^{eff}_{wk}|B>$ for the
helicity matrix element, $\lambda=0,\pm 1$. For decays of $B$ to
two vector mesons, the amplitude can be expressed by three
invariant helicity amplitudes, defined by the decomposition
\begin{equation}\label{pm}
M_{\lambda}=M^{(1)}\epsilon^*_{K^*}(\lambda)\cdot\epsilon^*_{\rho}(\lambda)
+M^{(2)}\epsilon^*_{K^*}(\lambda)\cdot P_{\rho}
\epsilon^*_{\rho}(\lambda)\cdot P_{K^*}
+M^{(3)}i\varepsilon_{\mu\nu\omega\sigma}\epsilon_{K^*}^{*\mu}(\lambda)
\epsilon_{\rho}^{*\nu}(\lambda)P^{\omega}_{K^*}P_{\rho}^{\sigma}.
\end{equation}
According to the naive counting rules mentioned before, we can
estimate that polarization fractions satisfy the relation:
$|M_0|^2 \gg |M_-|^2\gg |M_+|^2$. These three helicity amplitudes
can be expressed as another set of helicity amplitudes,
\begin{eqnarray}
M_0=M_B^2M_L,~M_\pm=M_B^2M_N {\mp}
M_{K^*}^2\sqrt{r^{\prime}-1}M_T
\end{eqnarray}
where the $ M_L$, $M_N$ and $M_T$ can be extracted directly from
calculation of the Feynman diagrams, and
$r^{\prime}=\frac{P_2\cdot P_3}{M_{K^*}M_{\rho}}$. The formula for
the decay width is
\begin{eqnarray}
\Gamma=\frac{p}{8\pi M_B^2}\sum
M_{(\sigma)}^{\dagger}M_{(\sigma)}.
\end{eqnarray}
Here $p$ is the absolute value of the 3-momentum of the final
state mesons. And we have
\begin{eqnarray}
\sum
M_{(\sigma)}^{\dagger}M_{(\sigma)}&&\nonumber=|M_0|^2+|M_+|^2+|M_-|^2
..
\end{eqnarray}

The weak Hamiltonian ${\cal H}_{{\it eff}}$ for the $\Delta B=1$
transitions at the scale smaller than $m_W$ is given as
\cite{buras}
\begin{equation}
\label{heff} {\cal H}_{{\it eff}} = \frac{G_{F}} {\sqrt{2}} \,
\left[ V_{ub} V_{us}^* \left (C_1 O_1^u + C_2 O_2^u \right) -
V_{tb} V_{ts}^* \, \left(\sum_{i=3}^{10} C_{i} \, O_i + C_g O_g
\right) \right] \quad .
\end{equation}
We specify below the operators in ${\cal H}_{{\it eff}}$ for $b
\to s$:
\begin{equation}\begin{array}{llllll}
 O_1^{u} & = & \bar s_\alpha\gamma^\mu L u_\beta\cdot \bar
u_\beta\gamma_\mu L b_\alpha\ , &O_2^{u} & = &\bar
s_\alpha\gamma^\mu L u_\alpha\cdot \bar
u_\beta\gamma_\mu L b_\beta\ , \\
O_3 & = & \bar s_\alpha\gamma^\mu L b_\alpha\cdot \sum_{q'}\bar
 q_\beta'\gamma_\mu L q_\beta'\ ,  &
O_4 & = & \bar s_\alpha\gamma^\mu L b_\beta\cdot \sum_{q'}\bar
q_\beta'\gamma_\mu L q_\alpha'\ , \\
O_5 & = & \bar s_\alpha\gamma^\mu L b_\alpha\cdot \sum_{q'}\bar
q_\beta'\gamma_\mu R q_\beta'\ ,  & O_6 & = & \bar
s_\alpha\gamma^\mu L b_\beta\cdot \sum_{q'}\bar
q_\beta'\gamma_\mu R q_\alpha'\ , \\
O_7 & = & \frac{3}{2}\bar s_\alpha\gamma^\mu L b_\alpha\cdot
\sum_{q'}e_{q'}\bar q_\beta'\gamma_\mu R q_\beta'\ ,  & O_8 & = &
\frac{3}{2}\bar s_\alpha\gamma^\mu L b_\beta\cdot
\sum_{q'}e_{q'}\bar q_\beta'\gamma_\mu R q_\alpha'\ , \\
O_9 & = & \frac{3}{2}\bar s_\alpha\gamma^\mu L b_\alpha\cdot
\sum_{q'}e_{q'}\bar q_\beta'\gamma_\mu L q_\beta'\ ,  & O_{10} & =
& \frac{3}{2}\bar s_\alpha\gamma^\mu L b_\beta\cdot
\sum_{q'}e_{q'}\bar q_\beta'\gamma_\mu L q_\alpha'~.
\label{operators}
\end{array}
\end{equation}
Here $\alpha$ and $\beta$ are the $SU(3)$ color indices; $L$ and
$R$ are the left- and right-handed projection operators with $L=(1
- \gamma_5)$, $R= (1 + \gamma_5)$. The sum over $q'$ runs over the
quark fields that are active at the scale $\mu=O(m_b)$, i.e.,
$(q'\epsilon\{u,d,s,c,b\})$.

\begin{figure}[[hbt]
\begin{center}
\epsfig{file=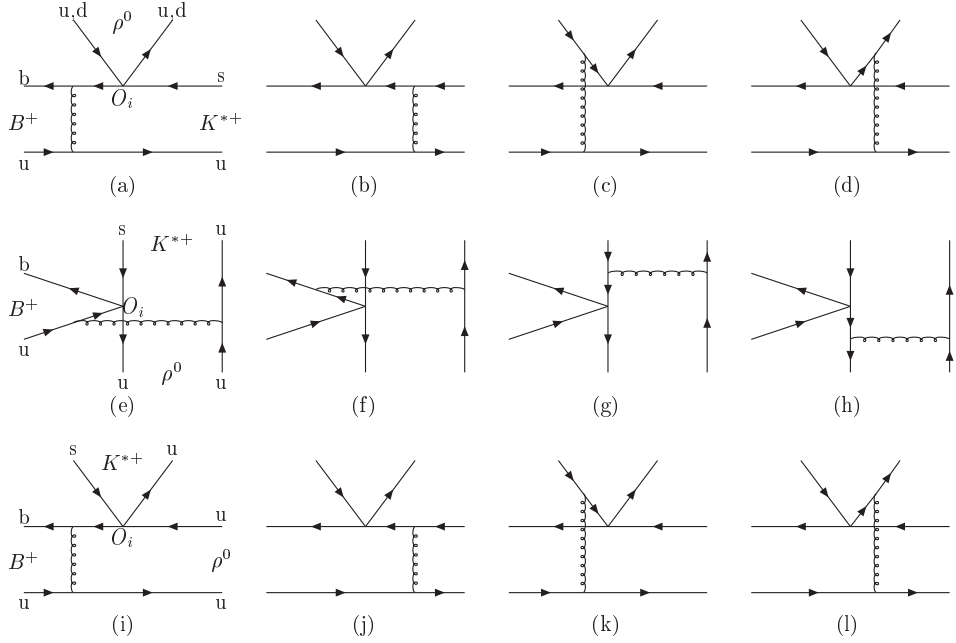,bb= 295 295 370 725,width=2.2cm}
\end{center}
\vspace{-3.5cm}
\caption{Diagrams contributing to the $B^+\to K^{*+}\rho^0$ decays}
\label{diagrams}
\end{figure}

The diagrams for these decays are completely the same as ones in
the decay $B\to K\pi$. Here we take the decay $B\to \rho^0 K^{*+
}$ as an example, whose diagrams are shown in Figure
\ref{diagrams}. These are all single hard gluon exchange diagrams,
containing all leading order PQCD contributions. The analytic
calculation is performed through the contraction of these hard
diagrams and the Lorenz structures of the mesons' wave functions.
The first row and the third row in Figure 1 are called emission
diagrams, with the $\rho$ meson or $K^*$ meson emitted. The
analytic formulae for the $K^*$ meson emission diagram is exactly
the same as the emission diagrams of $B\to K^{*+ }\phi$ with
$f_{\phi}\to f_{K^*},\,\,f_{K^*}\to f_{\rho}$, and we can get the
formulae for the $\rho$ emission diagrams through the change
$f_{\phi}\to f_{\rho},\,\,x_3 \to x_2$ from $B\to K^{*+ }\phi$. As
to the annihilation diagrams, we make the same change as the $K^*$
emission diagrams for the corresponding diagrams of $B\to K^{*+
}\phi$, then we can get the right analytic formulae.

In PQCD approach, only Wilson coefficients are channel dependent.
There are six different decay channels in $B^+(B^0)\to
\rho(\omega) K^*$ decays, and the $B^-\bar{(B^0)}$ decays are
their CP conjugation. All these decays are included in the twelve
diagrams, the only changes needed are external quarks and the
Wilson coefficients. We summarize the Wilson coefficients for each
channels in table 3.
 In this table the coefficients are defined as
\begin{eqnarray}
\nonumber &&a_1=C_1+C_2/N_c,~~~~~a_2=C_2+C_1/N_c, \\
\nonumber
&&a^q_3=C^q_3+C^q_4/N_c+\frac{3}{2}e_q(C_9+C_{10}/N_c),\\
\nonumber
&&a^q_4=C^q_4+C^q_3/N_c+\frac{3}{2}e_q(C_{10}+C_9/N_c),\\
\nonumber
&&a^q_5=C^q_5+C^q_6/N_c+\frac{3}{2}e_q(C_7+C_8/N_c),\\
&&a^q_6=C^q_6+C^q_5/N_c+\frac{3}{2}e_q(C_8+C_7/N_c),
\end{eqnarray}
and
\begin{eqnarray}
&&a^{\prime q}_3=C^q_3+\frac{3}{2}e_qC_9,\\
\nonumber
&&a^{\prime q}_4=C^q_4+\frac{3}{2}e_qC_{10},\\
\nonumber
&&a^{\prime q}_5=C^q_5+\frac{3}{2}e_qC_7,\\
&&a^{\prime q}_6=C^q_6+\frac{3}{2}e_qC_8.
\end{eqnarray}

\begin{table}
\begin{center}\caption{   Wilson coefficients
(the characters in the first row stand for the diagrams in figure
1 )}
\begin{tabular}{|l|l|l|l|l|l|l|}
\hline
 Process
  &(a)(b) &(c)(d)  &(g)(h) &(e)(f) &(i)(j) &(k)(l)
  \\ \hline
 $ B^0\to\rho^- K^{*+ }$&$$&$$
   &$a^{(d)}_4,a^{(d)}_6$&$a^{\prime(d)}_3,a^{\prime(d)}_5$&$a_2,a^{(u)}_4$&
   $C_1,a^{\prime(u)}_3,a^{\prime(u)}_5$\\
 \hline
 $ B^+\to\rho^0 K^{*+ }$&$a_1,a^{(u-d)}_3,a^{(u-d)}_5$&
 $C_2,a^{\prime(u-d)}_4,a^{\prime(u-d)}_6$
   &$a_2,a^{(u)}_4,a^{(u)}_6$&$C_1,a^{\prime(u)}_3,a^{\prime(u)}_5$&$a_2,a^{(u)}_4$&
   $C_1,a^{\prime(u)}_3,a^{\prime(u)}_5$
\\ \hline
 $ B^+\to\rho^+ K^{*0 }$ & $$ &  & $a_2,a^{(u)}_4,a^{(u)}_6$ &
 $C_1,a^{\prime(u)}_3,a^{\prime(u)}_5$
 & $a^{(d)}_4$ & $a^{\prime(d)}_3,a^{\prime(d)}_5$
\\ \hline
 $ B^0\to\rho^0 K^{*0 }$ &$a_1,a^{(u-d)}_3,a^{(u-d)}_5$
 &$C_2,a^{\prime(u-d)}_4,a^{\prime(u-d)}_6$
 & $a^{(d)}_4,a^{(d)}_6$
 & $a^{\prime(d)}_3,a^{\prime(d)}_5$
 & $a^{(d)}_4$ & $a^{\prime(d)}_3,a^{\prime(d)}_5$
\\ \hline
 $ B^0\to\omega K^{*0 }$ &$a_1,a^{(u+d)}_3,a^{(u+d)}_5$
 &$C_2,a^{\prime(u+d)}_4,a^{\prime(u+d)}_6$
 & $a^{(d)}_4,a^{(d)}_6$
 & $a^{\prime(d)}_3,a^{\prime(d)}_5$
 & $a^{(d)}_4$ & $a^{\prime(d)}_3,a^{\prime(d)}_5$
\\ \hline
 $ B^+\to\omega K^{*+ }$ &$a_1,a^{(u+d)}_3,a^{(u+d)}_5$&
 $C_2,a^{\prime(u+d)}_4,a^{\prime(u+d)}_6$
   &$a_2,a^{(u)}_4,a^{(u)}_6$& $C_1,a^{\prime(u)}_3,a^{\prime(u)}_5$&
   $a_2,a^{(u)}_4$& $C_1,a^{\prime(u)}_3,a^{\prime(u)}_5$
\\ \hline

\end{tabular}
\end{center}\end{table}

\section{Numerical Calculations and Discussions of Results}
\label{sc:NumCalc}

In the numerical calculations we use \cite{particle}
\begin{eqnarray}
  \nonumber
   f_B = 190 MeV, &
m_{K^*}=0.892 \mbox{ GeV},& m_\rho=0.77 \mbox{ GeV} , \nonumber\\
  M_B = 5.28 { GeV}, &f_{K^*} = 217
 { MeV}, &f_{K^*}^T = 160 MeV, \nonumber\\
   M_W = 80.41{ GeV},&f_{\rho} = 205
 { MeV}, &f_{\rho}^T = 155 MeV, \nonumber\\
m_\omega=0.782 \mbox{ GeV},  &  f_{\omega} = 195
 { MeV}, &f_{\omega}^T = 140 MeV,\nonumber\\
\tau_{B^\pm}=1.671\times 10^{-12}\mbox{ s,} &
\tau_{B^0}=1.536\times 10^{-12}\mbox{ s},
&\Lambda_{\overline{\mathrm{MS}}}^{(f=4)} = 250 { MeV}.
\end{eqnarray}
The distribution amplitudes $\phi_\rho^i(x)$ ($\phi_\omega^i(x)$) and
 $\phi_{K^*}^i(x)$ of the light mesons used in the numerical
 calculation are listed in Appendix A.

For $B$ meson, the wave function is chosen as
\begin{eqnarray}
\phi_B(x,b) &=& N_B x^2(1-x)^2 \exp \left
 [ -\frac{M_B^2\ x^2}{2 \omega_{b}^2} -\frac{1}{2} (\omega_{b} b)^2\right],
\label{phib}
\end{eqnarray}
with $\omega_{b}=0.4$ GeV  \cite{bsw}, and the normalization
constant $N_B=91.784$ GeV. We would like to point out that the
choice of the meson wave functions and the parameters above is the
result of a global fitting for $B\to \pi \pi$ and $B\to \pi K$
decays \cite{kls,lucd}.

For the CKM matrix elements, we use $| V_{us}  V_{ub}^*|
=0.00078$, $| V_{ts}  V_{tb}^*| = 0.0395$. We leave the CKM angle
$\phi_3$ as a free parameter, which is defined as
\begin{equation}
V_{ub}=|V_{ub}|exp(-i\phi_3).
\end{equation}
The decay amplitude of $B\to K^*\rho$ can be written as
\begin{eqnarray}
{\cal M}^{(i)} &=& V_{ub}^*V_{us} T^{(i)} -V_{tb}^* V_{ts} P^{(i)}\nonumber\\
 &=& V_{ub}^*V_{us} T^{(i)}
\left[1 -z^{(i)} e^{i(-\phi_3+\delta^{(i)})} \right],~~i=1,2,3,
\end{eqnarray}
where $z^{(i)}=\left|\frac{V_{tb}^* V_{ts}}{ V_{ub}^*V_{us} }
\right| \left|\frac{P^{(i)}}{T^{(i)}} \right|$, and $\delta^{(i)}$
is the relative strong phase between tree (T) diagrams and penguin
diagrams (P). $z^{(i)}$ and $\delta^{(i)}$ can be calculated
perturbatively. Here in PQCD approach, the strong phases come from
the non-factorizable diagrams and annihilation type diagrams (see
(c) $\sim$ (h) in Figure \ref{diagrams}). The internal quarks and
gluons can be on mass shell, and then poles appear in the
propagators, which can provide the strong phases. The predominant
contribution to the relative strong phase $\delta$ comes from the
annihilation diagrams, (g) and (h) in Figure \ref{diagrams}.

This mechanism of producing   strong phase is very different from
the so-called Bander-Silverman-Soni (BSS) mechanism
 \cite{bss}, where the strong phase comes from the perturbative charm penguin
diagrams. The contribution of BSS mechanism to the direct CP
violation in $B\to K^*\rho$ is only in the higher order corrections
($\alpha_s$ suppressed) in our PQCD approach. Therefore we can safely
neglect this contribution.

\begin{figure}[[hbt]
\begin{center}
\psfig{file=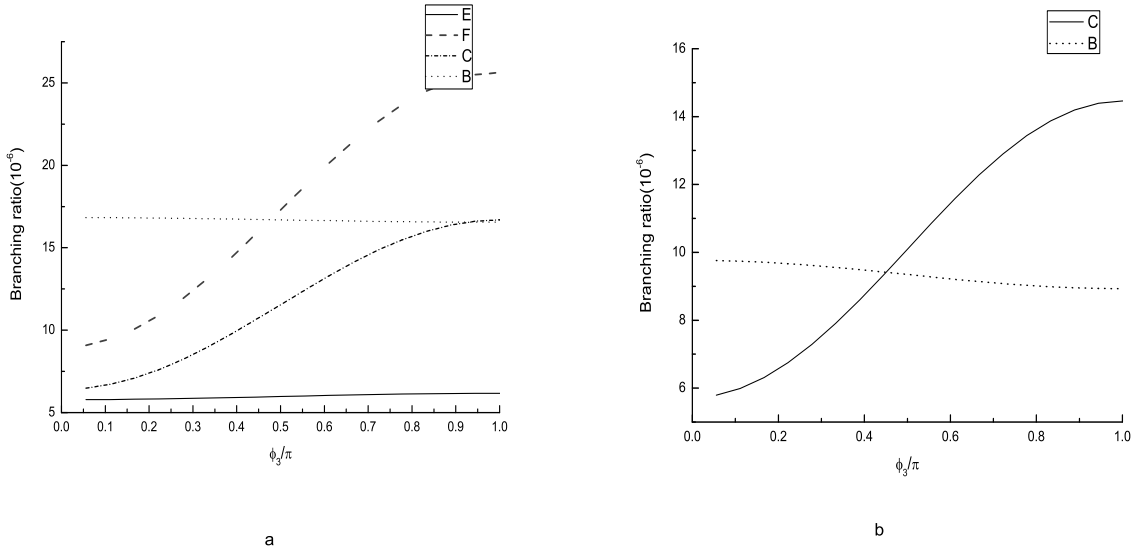,width=18.7cm,angle=0}
\end{center}
\vspace{-1cm} \caption{Averaged branching ratios ($10^{-6}$) of
$B\to K^*\rho$ and $B\to K^*\omega$ as a function of CKM angle
$\phi_3$, where the lines B,C,E,F in diagram (a) represent $B^+\to
\rho^+ K^{*0}$,$B^+\to \rho^0 K^{*+}$, $B^0\to \rho^0 K^{*0}$,
$B^0\to \rho^- K^{*+}$ respectively, and in diagram (b), B denotes
$B^0\to \omega K^{*0}$, C denotes $B^+\to \omega K^{*+}.$}
\label{bran}
\end{figure}

The corresponding charge conjugate $\bar B$ decay is
\begin{eqnarray}
{\cal M}^{(i)} &=& V_{ub}V_{us}^* T^{(i)} -V_{tb} V_{ts}^* P^{(i)}\nonumber\\
 &=& V_{ub}V_{us}^* T^{(i)}
\left[1 -z^{(i)} e^{i(\phi_3+\delta^{(i)})} \right].
\end{eqnarray}
In contrast to the decay of $B$ to pseudoscalar mesons like $B\to
K\pi$, where the decay widths can be expressed in terms of
$\delta$ and $\phi_3$ in a simple way, here for $B$ decay to two
vector mesons, there are 3 types of amplitudes, and this makes the
dependence of decay widths on $\delta$ and $\phi_3$ very
complicated.
 The averaged
decay width for $B$ and its CP conjugation decays can be expressed
as a function of a CKM phase angle $\phi_3$.
\begin{eqnarray}
\Gamma=\frac{ p}{8\pi M_B^2}
|V_{ub}^*V_{us}|^2[T_L^2(1+z_L^2+2z_Lcos\phi_3 \cos
\delta_L)+2\sum_{i=N,T}T_i^2(1+z_i^2+2z_icos\phi_3 \cos
\delta_i)].
\end{eqnarray}
 From this formula we can know that when contribution from the
penguin diagrams  are much larger than that from the tree
diagrams, i.e., $z_i\gg 1 , i= L,N,T$, then the branching ratios
are insensitive to the angle, but when they are comparable, the
dependence on $\phi_3$ will be strong.
 We show the branching ratios of these decays in Figure 2, from which we
 can see that  the penguin dominant decays $B^+\to \rho^+
K^{*0}$,$B^0\to \rho^0 K^{*0}$, and $B^0\to \omega K^{*0}$ are
almost independent on $\phi_3$, but the dependence on $\phi_3$ of
the other three channels is strong, because of the tree and
penguin interference.

\begin{figure}[[hbt]
\begin{center}
\psfig{file=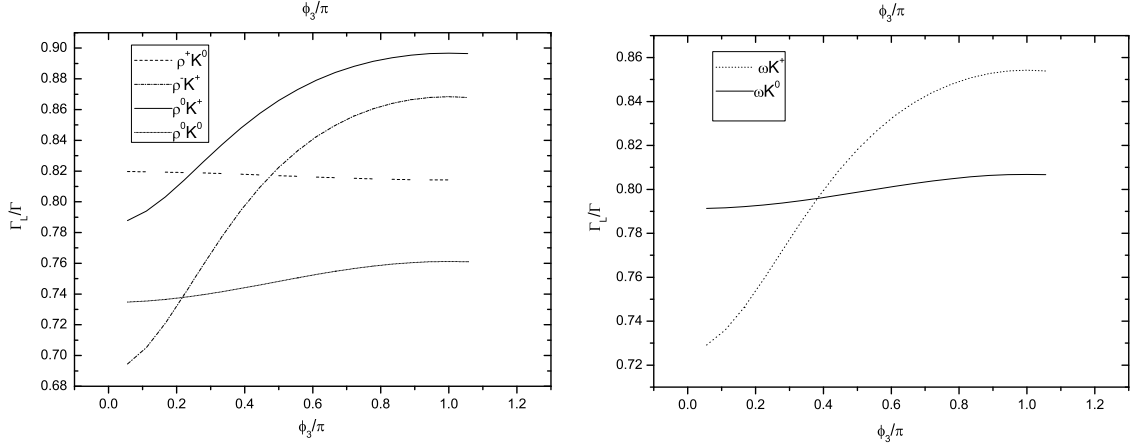,width=18.7cm,angle=0}
\end{center}
\vspace{-1cm}
\caption{Longitudinal polarization fraction of $B\to
K^*\rho$ and $B\to K^*\omega$ as a function of CKM angle $\phi_3$}
\label{gt}
\end{figure}

In Figure 3, we  plot the dependence of longitudinal polarization
fractions $\Gamma_L/\Gamma$ on the CKM angle $\phi_3$.  We find
that this quantity is not very sensitive to $\phi_3$ in all decay
channels. If we fix $\phi_3$ at about $60^{\circ}$, we find that
for the decays $B^+\to K^{*+}\rho^0$ and $B^+\to K^{*0}\rho^+$,
the longitudinal fractions are 0.89 and 0.82 respectively. As
mentioned before, we calculate the annihilation type diagrams in
PQCD approach. If the four quark operator has the Dirac structure
like $(S-P)(S+P)$, there is no helicity flip suppression   to the
transverse polarization, so that the longitudinal fractions are
considerably suppressed. One can see that our results for $B^+\to
K^{*0}\rho^+ $ are consistent with   BaBar, but different from
Belle (we hope more efforts from experimental side to test our
prediction). As to $B^+\to K^{*+}\rho^0 $, our result is a little
smaller, but still agree with the data within the 1$\sigma$ error
bar.

The new analysis of the $K^*$ meson wave function from QCD sum
rules
 \cite{braun} shows that the leading twist distribution amplitude $\phi_{K^*}(x)$
 of longitudinal polarization should be very close to
the asymptotic one. According to Li's suggestion \cite{reso}, we
test our result using the asymptotic wave functions for the
longitudinal polarization part. The numerical results are given in
table 4. We find that the longitudinal fraction  and the branching
ratios for all the channels are reduced. Note that Figure~2 shows
that the branching ratios of $B^o\to K^{*0}\omega $ and $B^+\to
K^{*+}\omega $ are larger than the experimental limits where we
use the wave functions given in the appendix. But if we adopt the
asymptotic form,   the branching ratios decrease. Comparing the
table with the experimental data, it seems that the asymptotic
form is more convincing. More   study of the vector meson's wave
functions are required.

\begin{table}\begin{center}\caption{ Branching ratios ($10^{-6}$) and
polarization fractions using different type of light meson wave
functions (the CKM phase angle $\phi_3$ is fixed as $60^{\circ}$;
the w.f. stands for $wave\, function$)}

\begin{tabular}{|l|l|l|l|l|l|}
\hline
 Quantity &  w.f. in the appendix &  asymptotic w.f.
  \\ \hline\hline
$Br(B^0\to\rho^- K^{*+ })$&13&9.8
 \\ \hline
 $ Br(B^+\to\rho^+ K^{*0 })$&$17$& $13$
\\ \hline

$ Br(B^+\to\rho^0 K^{*+ })$&9.0&6.4
\\ \hline
 $ Br(B^0\to\rho^0 K^{*0 })$& $5.9 $ & $4.7$
\\ \hline
 $ Br( B^+\to\omega K^{*+ })$&7.9&5.5
\\ \hline
 $ Br(B^0\to\omega K^{*0 })$&$9.6 $& $6.6$
\\ \hline\hline

$R_L(B^0\to\rho^- K^{*+ })$&$0.78$& $0.71$
 \\ \hline
 $ R_L(B^+\to\rho^+ K^{*0 })$&$0.82$& $0.76$
\\ \hline
 $ R_L(B^+\to\rho^0 K^{*+ })$&$0.85$& $0.78$

\\ \hline
 $ R_L( B^+\to\omega K^{*+ })$ & $0.81$& $0.73$
 \\ \hline
 $ R_L(B^0\to\rho^0 K^{*0 })$&$0.74$& $0.68$
 \\ \hline
 $ R_L(B^0\to\omega K^{*0 })$&$0.82$& $0.74$

\\ \hline
\end{tabular}
\end{center}\end{table}

\begin{figure}[[hbt]
\begin{center}
\psfig{file=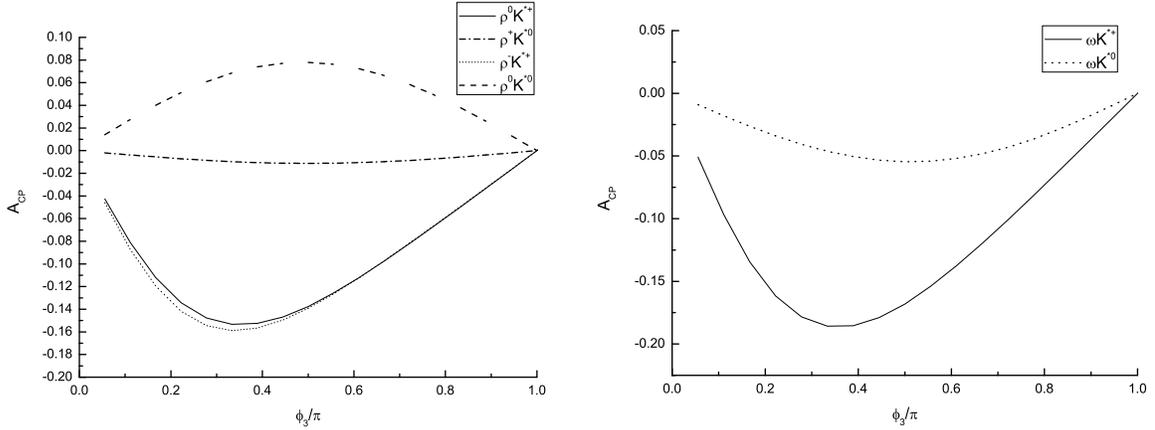,width=18.7cm,angle=0}
\end{center}
\vspace{-1cm} \caption{Direct CP violation of $B\to K^*\rho$ and
$B\to K^*\omega$ as a function of CKM angle $\phi_3$} \label{cp}
\end{figure}

It has been confirmed that there is big direct CP violation in
$B\to \pi K$ and $B\to \pi \pi$ decays \cite{hfag}, and the PQCD
approach can give right predictions from the annihilation topology
\cite{direct} rather than the BSS mechanism. Here we take the
definition (note that our definition has opposite sign when
comparing with the definition used in \cite{hfag})
\begin{eqnarray}
A_{CP}=\frac{\Gamma(B\to f)-\Gamma(\bar{B}\to
\bar{f})}{\Gamma(B\to f)+\Gamma(\bar{B}\to \bar{f})}.
\end{eqnarray}
The direct CP violation parameters   as a function of $\phi_3$ are
shown in Figure~\ref{cp}. Since CP asymmetry is sensitive to many
parameters, the line should be broadened by uncertainties. The
direct CP violation parameter of $B^+\to K^{*+}\rho^0$, $B^0\to
K^{*+}\rho^-$ and $B^+\to K^{*+}\omega$ can be large as $15-20\%$
when $\phi_3$ is near $60^\circ$, but for $B^+\to K^{*0}\rho^+$,
the direct CP violation is very small for the very tiny tree
diagram contribution.  The final state is not the CP eigenstate,
so the mixing induced CP violation is more complicated, and we do
not give it here.

The angular distributions depend on the spins of the decay
products of the decay vector mesons $K^*$ and $\rho$. For example,
for $B^+\to K^{*+}\rho^0\to (K\pi)(\pi^+\pi^-)$ the differential
decay distribution is  \cite{palmer} \bea\label{angle}
\frac{d^3\Gamma}{d\cos\theta_1d\cos\theta_2d\phi}
&=&\frac{9}{8\pi}\Gamma\left\{ \frac{1}{4}\frac{\G_T}{\G} \cdot
\sin^2\theta_1\sin^2\theta_2 +\frac{\G_L}{\G} \cdot
\cos^2\theta_1\cos^2\theta_2\right.
\nonumber \\[3mm]
&&+\frac{1}{4}\sin2\theta_1 \sin2\theta_2 \lbrack \a_1 \cdot
\cos\phi-\b_1 \cdot \sin\phi\rbrack
\label{distribution} \nonumber \\[3mm]
&&\left.+\frac{1}{2}\sin^2\theta_1 \sin^2\theta_2\ \lbrack \a_2
\cdot \cos2\phi\ - \b_2 \cdot \sin2\phi\ \rbrack\right\} \ . \eea
In (\ref{distribution}) $\theta_1$ is the polar angle of the $K$
in the rest system of the $K^*$ with respect to the helicity axis.
Similarly $\theta_2$ is the polar angle of the $\pi^+$ in the
$\rho^0$ rest system with respect to the helicity axis of the
$\rho^0$, and $\phi$ is the angle between the planes of the two
decays $K^{*-}\to K\pi$ and $\rho^0 \to \pi^+\pi^-$. The
coefficients in the decay distribution are related to the helicity
matrix elements by \be\label{xishu}\ba{llllll}
\frac{\Gamma_T}{\Gamma} & = & \frac{\vert M_{+1}\vert ^2 + \vert
M_{-1}\vert ^2} {\vert M_0\vert ^2 + \vert M_{+1}\vert ^2 + \vert
M_{-1}\vert ^2}, \ \ \ \ \ & \frac{\Gamma_L}{ \Gamma} & = &
\frac{\vert M_0\vert ^2}{\vert M_0\vert ^2 + \vert M_{+1}\vert ^2
+\vert M_{-1}
\vert ^2}, \\[5mm]
\alpha_1 & = & \frac{Re\left (M_{+1}M_0^\ast +
M_{-1}M_0^\ast\right )} {\vert M_0\vert ^2 + \vert M_{+1}\vert ^2
+ \vert M_{-1}\vert ^2}, & \beta_1 & = &\frac{Im\left (M_{+1}
M_0^\ast- M_{-1}M_0^\ast\right)}
{\vert M_0 \vert ^2+ \vert M_{+1} \vert ^2 + \vert M_{-1} \vert ^2 },\\[5mm]
\alpha_2 & = & \frac{Re\left(M_{+1}M_{-1}^\ast\right)} {\vert
M_0\vert ^2 + \vert  M_{+1}\vert ^2 + \vert M_{-1}\vert ^2}, &
\beta_2 & = & \frac{Im\left (M_{+1} M_{-1}^\ast\right)}
{\vert M_0\vert ^2 +\vert M_{+1}\vert ^2 + \vert M_{-1}\vert ^2}. \\
\ea \label{parameters}\ee

\vspace{2mm}
 The integration over angles $\theta_1$, $\theta_2$ in
(\ref{angle}) yields the $\phi$ distribution of the decay width
\begin{eqnarray}\label{jaodu}
2\pi\frac{d\Gamma}{d\phi}&=&\Gamma(1 +2\alpha_2\cos
2\phi-2\beta_2\sin2\phi),
\end{eqnarray}
where the coefficients $\alpha_2$, $\beta_2$  can be obtained from
(\ref{xishu}) by using the $M_\lambda$ which is calculated in PQCD
approach. Because  $\alpha_2$ and $\beta_2$ are very small, the
decay width is almost independent in $\phi$, then the CP violation
 from the angular distribution will be very tiny in the standard
model.

\section{Summary}

We performed the calculations of $B^+\to K^{*+}\rho^0$, $B^+\to
K^{*0}\rho^+$, $B^0\to K^{*+}\rho^-$, $B^0\to K^{*0}\rho^0$ and
$B^+\to K^{*+}\omega$, $B^0\to K^{*0}\omega$ in PQCD approach. In
this approach, we calculated the non-factorizable contributions and
annihilation type contributions in addition to the usual factorizable
contributions.

We found that the annihilation contributions were not so small as
expected in a simple argument. The annihilation diagram, which
provides the dominant strong phases, plays an important role in
the direct CP violations. We expect large direct CP asymmetry in
the decays of $B^+\to K^{*+}\omega$, $B^0\to K^{*+}\rho^-$ and
$B^+\to K^{*+}\rho^0$. We also study the helicity structure and
angular distribution of the decay products. The current running B
factories in KEK and SLAC will be able to test the theory.

\section*{Acknowledgments}

This work started when one of the authors (H.W.H.) was at Kobe
University, where he was supported by the JSPS (Grant No. P99221).
This work is partly supported by the National Science Foundation
of China under Grant No.90103013, 10475085  and 10135060. We thank
H.n. Li for helpful discussions.

\begin{appendix}

\section{Wave Functions of Light Mesons Used in the Numerical Calculation}

 For the light meson wave function, we neglect the $b$ dependence
 part, which is not important in numerical analysis.
 We choose the different distribution amplitudes of ${\rho}$ meson longitudinal
 wave function as  \cite{ball2},
\begin{eqnarray}
\phi_{\rho}(x) &=& 6f_{\rho}
  x (1-x) \left[1+ 0.18C_2^{3/2} (t) \right],
 \\
  \phi_{\rho}^t(x) &=& f_{\rho}^T
 \left\{ 3 t^2 +0.3t^2 \left[5t^2-3 \right]
 +0.21 \left[3- 30 t^2 +35 t^4\right] \right\},\\
\phi_{\rho}^s(x) &=& 3f_{\rho}^T
 ~ t \left[1+ 0.76 (10 x^2 -10 x +1) \right],
\end{eqnarray}

where $t=1-2x$. The Gegenbauer polynomials are defined by
\begin{equation}
\begin{array}{ll}
 C_2^{1/2} (t) = \frac{1}{2} (3t^2-1), & C_4^{1/2} (t) = \frac{1}{8}
 (35t^4-30t^2+3),\\ C_2^{3/2} (t) = \frac{3}{2} (5t^2-1), & C_4^{3/2}
 (t) = \frac{15}{8} (21t^4-14t^2+1).
\end{array}
\end{equation}

  For the transverse $\rho$ meson we use  \cite{ball2}:
  \begin{eqnarray}
\phi_{\rho}^T(x) &=& 6f_{\rho}^T
 x (1-x) \left[1+ 0.2C_2^{3/2} (t) \right],
 \\
  \phi_{\rho}^v(x) &=& f_{\rho}
 \left\{ \frac{3}{4} ( 1+ t^2) +0.24( 3 t^2-1) +0.12 ( 3-30 t^2
 +35 t^4) \right\},\\
\phi_{\rho}^a(x) &=& \frac{3 f_{\rho}}{2
 }~ t \left[1+ 0.93 (10 x^2 -10 x +1) \right] .
\end{eqnarray}
 For the $\omega$ meson, we use the same as the above $\rho$
 meson, except changing the decay constant $f_\rho$ with $f_\omega$.

 We choose the light cone distribution amplitudes of ${K^*}$ meson longitudinal wave
  function as  \cite{ball2},
\begin{eqnarray}
\phi_{K^*}(x) &=& 6 f_{K^*}
 x (1-x) \left[1+ 0.57t + 0.07C_2^{3/2} (t) \right],
 \\
  \phi_{K^*}^t(x) &=& f_{K^*}^T
 \left\{ 0.3 t (3t^2 +10 t-1) +1.68 C_4^{1/2} (t)
 +0.06t^2(5t^2-3)  \right.  \nonumber\\
&& \left. +0.36 [1 -2 t -2t \ln(1- x) ] \right\},
\\
\phi_{K^*}^s(x) &=& f_{K^*}^T
\left\{ 3 t \left[1+ 0.2t+0.6 (10 x^2 -10 x +1)\right] -0.12x(1-x)   \right.  \nonumber\\
&& \left. +0.36 [1 -6x-2 \ln(1- x)] \right\}.
\end{eqnarray}
The light cone distribution amplitudes of $K^*$ transverse  wave
function are used as \cite{ball2}
  \begin{eqnarray}
\phi_{K^*}^T(x) &=& 6 f_{K^*}^T
 x (1-x) \left[1+ 0.6t+0.04C_2^{3/2} (t) \right],
 \\
  \phi_{K^*}^v(x)
     &=& f_{K^*}
   \left\{ \frac{3}{4} (1+t^2 +0.44 t^3) +0.4 C_2^{1/2} (t)
   +0.88C_4^{1/2}(t)   \right.  \nonumber\\
  && \left. +0.48 [2x + \ln(1- x)] \right\},
  \\
\phi_{K^*}^a(x)
   &=& \frac{f_{K^*} }{2 }
  \left\{ 3 t \left[1+ 0.19t+0.81 (10 x^2 -10 x +1)\right] -1.14 x(1-x)   \right.
    \nonumber\\
  && \left. +0.48 [1 -6x-2 \ln(1- x)] \right\}.
\end{eqnarray}

\end{appendix}

\end{document}